\documentclass[]{iopart}

\usepackage{graphicx}

\usepackage{nicefrac}

\usepackage{ulem}
\usepackage[dvipsnames,usenames]{color} 

\DeclareTextFontCommand{\emph}{\itshape}
\def\rmd{\mathrm{d}}

\begin{document}

\title[CEP-stable Tunable THz-Emission from Controlled Plasma-Electron Bursts]{CEP-stable Tunable THz-Emission Originating from Laser-Waveform-Controlled Sub-Cycle Plasma-Electron Bursts}

\author{T. Bal\v{c}i\={u}nas$^1$, D. Lorenc$^1$, M. Ivanov$^{2}$, O. Smirnova$^{2}$, A. M. Zheltikov$^3$, D. Dietze$^1$, K. Unterrainer$^1$, T. Rathje$^4$, G. Paulus$^4$, A. Baltu\v{s}ka$^1$, S.Haessler$^1$\footnote{Present address: Laboratoire d'Optique Appliqu\'ee, ENSTA-Paristech -- Ecole Polytechnique -- CNRS, 91762 Palaiseau, France.}}

\address{1 Photonics Institute, Vienna University of Technology, Gu\ss hausstra\ss e 27/387, 1040 Vienna, Austria.}
\address{2 Max Born Institute, Max-Born-Stra\ss e 2 A, 12489 Berlin, Germany.}
\address{3 Department of Physics, International Laser Center, M.V. Lomonosov Moscow State University, Vorob'evy Gory, Moscow 119992, Russia}
\address{4 Institute of Optics and Quantum Electronics, Friedrich-Schiller-Universit\"at, Max-Wien-Platz 1, 07743 Jena, Germany.}
\ead{tadas.balciunas@tuwien.ac.at}

\date{\today}

\begin{abstract}
We study THz-emission from a plasma driven by an incommensurate-frequency two-colour laser field. A semi-classical transient electron current model is derived from a fully quantum-mechanical description of the emission process in terms of sub-cycle field-ionization followed by continuum-continuum electron transitions. For the experiment, a CEP-locked laser and a near-degenerate optical parametric amplifier are used to produce two-colour pulses that consist of the fundamental and its near-half frequency. By choosing two incommensurate frequencies, the frequency of the CEP-stable THz-emission can be continuously tuned into the mid-IR range. This measured frequency dependence of the THz-emission is found to be consistent with the semi-classical transient electron current model, similar to the Brunel mechanism of harmonic generation.
\end{abstract}

\maketitle

\section{Introduction}
Since the demonstration a decade earlier by Cook and Hochstrasser \cite{Cook:00} of intense THz emission from air ionized with a two-colour ($\omega+2\omega$) laser field, this phenomenon remains in the focus of attention. This is motivated by its potential as a source of intense single-cycle THz pulses, enabled by the high generated THz-field strength and large bandwidth and facilitated by the absence of material damage. The THz-range covers a wealth of fundamental resonances (in molecules: vibrational and rotational resonances, in solids: phonon and plasmon resonances, impurity transitions), which opens a wide field of possibilities for fundamental material and device research~\cite{Kampfrath2013resonant,Vicario2013magnetization} as well as possible sensor applications, including the identification of atmospheric pollutants and use in food quality-control, atmospheric and astrophysical remote sensing, and (medical) imaging with unique contrast mechanisms~\cite{Tonouchi2007THzreview}.  

Another motivation is the the interesting underlying plasma dynamics leading to the THz emission. Both an $\omega - (2\omega - \omega)$ four-wave mixing mechanism based on stationary or nonstationary third-order susceptibilities~\cite{Houard:08}, as well as a tunnel-ionization-induced micro-current mechanism~\cite{ISI:000259649700010,ISI:000237648400021} have been considered. According to the electron current model, the THz emission from plasma originates from sharp quasi-periodic changes of free electron density occurring at the peaks of optical field ionization in the tunneling regime. The induced electron current rapidly varies on the sub-cycle time scale and yields generation of many harmonics, also known as Brunel harmonics \cite{Brunel:90,PhysRevLett.104.163904}. Similarly to HHG where the mixing of the laser driving field with its second harmonic allows breaking the symmetry between the two quantum paths and yields generation of even harmonics \cite{ISI:000242477600018}, the THz emission can be considered an even ($0^{\mathrm{th}}$-order) harmonic. This particular mechanism of THz emission is thus closely related to two prominent phenomena, Above Threshold Ionization (ATI) and High-order Harmonic Generation (HHG), also based on quasi-periodic sub-cycle ionization followed by electron acceleration in the driving laser field. High-order harmonics are generated via transitions of the electron between a continuum state and the ground state, which can be efficiently controlled on the single-atom and sub-cycle level~\cite{PhysRevLett.102.063003,Haessler2014perfectwave}.

Here we show that the low-frequency THz emission corresponds to transitions between continuum states and can also be controlled by shaping the laser field. From quantum-mechanical continuum-continuum transition matrix elements, we recover, using the Strong Field Approximation (SFA) with a stationary phase analysis, the semi-classical interpretation in terms of transient electron currents by Brunel \cite{Brunel:90} and Kim et al. \cite{ISI:000259649700010}. 
Furthermore, we experimentally demonstrate control over the timing of sub-cycle optical field ionization relative to the sign and value of the vector potential of the pulse~\cite{Babushkin2011Tailoring}. This is achieved by by fixing the carrier envelope phases (CEP) and detuning the frequencies of the two optical driving fields away from commensurability~\cite{Thomson2010THzwhitelight}. This results in a very wide continuous tunability of the THz frequency emitted by the laser-generated plasma.

This frequency dependence of THz emission received little attention as mostly the commensurate $\omega+2\omega$ schemes have been used so far. Incommensurate frequency combinations have been realized by superimposing a broadband 20-fs pulse with its narrowband second harmonic; however the tunability range in this case is limited to the bandwidth of the fundamental~\cite{Thomson2010THzwhitelight}. Employing the relatively broad tuning range of the OPA, we obtained the initial evidence for the wide frequency-tunability of the THz-emission~\cite{up2010_thz:10,Balciunas201392}. Also, a scheme similar to our experimental setup based on a near-degenerate optical parametric amplifier (OPA) superimposed with the fundamental laser pulse has been implemented~\cite{Vvedenski2014halfharmonic}. However, due to lack of CEP stabilisation of the driving laser, field-resolved detection of the generated THz pulse was impossible. Furthermore, in that work only the lowest THz-frequencies ($<3\:$THz) and only very small detunings from the exact degenerate case have been considered.

Here, we demonstrate an incommensurate-two-colour scheme for THz generation, where the ratio $\omega_{1}/\omega_{2}$ is not an integer, based on a nearly degenerate OPA pumped by a CEP-stable Ytterbium-based laser emitting at 1030-nm wavelength. Inspired by our initial experiments \cite{up2010_thz:10,Balciunas201392}, we present extensive experimental data, theory and calculations that agree with the measurement data. We analyse the spectral properties of the generated  THz emission over a wide range of detuning frequencies from the exact commensurate case. With the CEP stabilised input pulses we are able to generated CEP-locked THz pulses and detect them in a field-resolved measurement, thus demonstrating their applicability for time-domain spectroscopy probing the amplitude and phase response of a material. Due to the long wavelengths of the constituent colours (1030~nm and 1800--2060~nm), our scheme is a promising route for a high-efficiency~\cite{Clerici2013wavelengthscaling}, CEP-locked, broadband and widely tunable mid-IR/THz source.

\begin{figure}
\begin{centering}
\includegraphics[width=.9\textwidth]{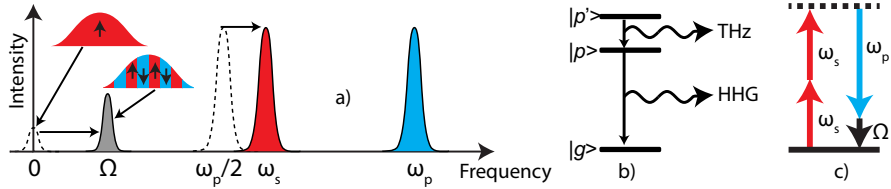}
\par\end{centering}

\caption{(a) Scheme of mixing two incommensurate frequencies for generating tunable THz
emission. (b) Scheme of the atomic transitions in a strong laser field: transitions between a continuum and a ground
state yield HHG, while transitions between continuum states yield low order harmonic generation (including THz). (c) Perturbative four-wave mixing scheme for THz generation. \label{fig:tuning_scheme}}
\end{figure}

\section{Quantum mechanical transient electron-current model for THz emission}
A proper quantum mechanical treatment of THz emission is desirable to support the commonly used semi-classical two-step model by Brunel \cite{Brunel:90} and Kim et al. \cite{ISI:000259649700010}. A very successful fully quantum mechanical model that describes emission of an atom exposed to a strong laser field is based on the SFA \cite{PhysRevA.49.2117}, where the influence of the atomic potential on the electron in the continuum, as well as the influence of all bound states except the ground state are neglected. Recently full ab-initio quantum TDSE calculations were performed to calculate THz emission \cite{PhysRevLett.102.093001,PhysRevA.79.063413}, but the drawback is that the physical nature  of the underlying dynamics is difficult to interpret. Here we use an analytical SFA model to calculate the atomic emission and derive the expressions that are identical to those previously attained in a classical description by Brunel of recollision-free harmonic emission.

An electron in a continuum interacting with the strong laser field can be described as a wavepacket, which propagates in phase space:

\begin{equation}
\Psi_\mathrm{c}(t)=-i\int_{0}^{t}\hat{U}(t,t_\mathrm{b})\hat{V}_{L}(t_\mathrm{b})\hat{U}_{0}(t_\mathrm{b},0)\Psi_{0} \,\rmd t',
\label{eq:wavepacket_propagation}
\end{equation}
%
{where $\hat{U}_{0}(t_1,t_0)$ and $\hat{U}(t_1,t_0)$ are the field-free and
full propagators between times $t_0$ and $t_1$ and the operator $V_{L}(t)$ describes the interaction
with the laser field. Here the velocity gauge is used. The interpretation of Eq.~(\ref{eq:wavepacket_propagation}),
which has proven both very fruitful and accurate, is that $t_\mathrm{b}$
are the instants of strong-field ionization and $\Psi_\mathrm{c}(t)$ is a
superposition of many wavepackets emitted into continuum at times
$t_\mathrm{b}$. Calculations based on semi-classical wavepacket propagators
or quantum-trajectories explicitly utilize this picture by representing
quantum mechanical amplitudes as sums over contributions of wavepackets
moving along classical trajectories \cite{PhysRevA.77.033407}. Therefore,
we express Eq.~(\ref{eq:wavepacket_propagation}) as a sum over such
wavepackets $\psi(r(t,t_\mathrm{b}),t)$, emitted by the parent neutral system
at times $t_\mathrm{b}$ and propagated to $t$:
\begin{equation}
\Psi_\mathrm{c}=\sum_{t_\mathrm{b}}\psi_{t_\mathrm{b}}\left[r(t,t_\mathrm{b}),t\right].
\label{eq:sum_of_wavepackets}
\end{equation}
Note that we formulate these equations for a linearly polarized laser field, so the electron dynamics are limited to a single dimension along that polarization direction.

The emission of electromagnetic radiation is proportional to the dipole acceleration, i.e. the time-derivative of the dipole
velocity} $\dot{d}_\mathrm{cc}(t)\equiv\left\langle \Psi_\mathrm{c}\left|\hat{p}\right|\Psi_\mathrm{c}\right\rangle$,
where $\hat{p}=i\nabla$ is the momentum operator (atomic units are used throughout the paper).
With (\ref{eq:sum_of_wavepackets}), the dipole velocity of the whole wavepacket is given by the sum of dipole velocities for all birth times: $\dot{d}_\mathrm{cc}(t) = \sum_{t_\mathrm{b1},t{}_\mathrm{b2}}\left\langle \psi_{t_\mathrm{b2}}\left|\hat{p}\right|\psi_{t_\mathrm{b1}}\right\rangle$. 
Expressing the wavefunction of the free electron born at instant $t_\mathrm{b}$ as a plane wave, $\psi_{t_\mathrm{b}}\left[r(t,t_\mathrm{b}),t\right]=\sqrt{\rho\left(t_\mathrm{b},t,r\right)}e^{iS(t_\mathrm{b},t,r)}$,
where $S=p r-E\left(t-t_\mathrm{b}\right)$ is the action and $\rho$ is the electron density, it can be shown (see Appendix), that the diagonal terms ($t_\mathrm{b1}=t_\mathrm{b2}$ ) are the stationary points in phase space and thus dominate this sum. 
Wavepackets emitted at different times, $t_\mathrm{b1}\neq t_\mathrm{b2}$, are separated in space, i.e. have a reduced overlap and lead to rapid phase oscillations.
%
%
We can thus simplify the dipole velocity to a single sum over birth instants:
\begin{eqnarray}
\dot{d}_\mathrm{cc}(t) & = & \sum_{t_\mathrm{b}}\left\langle \psi_{t_\mathrm{b}}\left|\hat{p}\right|\psi_{t_\mathrm{b}}\right\rangle, \label{eq:whole_momentum}\end{eqnarray}
and express the continuum-continuum transition dipole velocity matrix elements as:
\begin{equation}
\left\langle \psi_{t_\mathrm{b}}\left|\hat{p}\right|\psi_{t_\mathrm{b}}\right\rangle =\int \rmd r\rho\left(t_\mathrm{b},t,r\right) v\left(r,t,t_\mathrm{b}\right)+\delta\dot{d}(t)\label{eq:two_parts_matrix_element},\end{equation}
with the velocity $v=\partial S/\partial r$. Here, the first term describes the case when the wavepacket overlaps
with itself and $\delta\dot{d}(t)$ provides corrections. These may arise if some wavepackets emitted at different times have met at some
place in the phase space later.

This is an important difference as compared to HHG, where the continuum
and bound parts of the two wavepackets evolve differently and accumulate
a phase difference that depends on frequency and ultimately leads to
chirped emission of attosecond pulses. In C-C transitions, however,
the requirement for the wavepackets to overlap yields the Brunel harmonic
radiation which is unchirped.

We now concentrate on the first term in Eq. (\ref{eq:two_parts_matrix_element}) and rewrite:
\begin{equation}
\hspace{-15mm}\dot{d}_\mathrm{cc}(t)=\sum_{t_\mathrm{b}}\underbrace{\int\rho\left(t_\mathrm{b},t,r\right)\rmd r}_{W(t,t_\mathrm{b})} \,\underbrace{\frac{\int\rho\left(t_\mathrm{b},t,r\right)v\left(r,t_\mathrm{b},t\right)\rmd r}{\int\rho\left(t_\mathrm{b},t,r\right)\rmd r}}_{\overline{v}(t,t_\mathrm{b})} =\sum_{t_\mathrm{b}}W(t,t_\mathrm{b})\overline{v}(t,t_\mathrm{b})\end{equation}
as a sum over all ionization instants $t_\mathrm{b}$ of the product of $W(t,t_\mathrm{b})$, the norm of the wavepacket created at time $t_\mathrm{b}$,
and $\overline{v}(t,t_\mathrm{b})$, the average velocity of the wavepacket.

The wavepacket norm is time-independent and is given by the ionization
probability at $t_\mathrm{b}$: $W(t,t_\mathrm{b})=\Gamma(t_\mathrm{b})\delta t_\mathrm{b}$, where
$\delta t_\mathrm{b}$ is the saddle point region and $\Gamma$ the instantaneous ionization rate. Thus we have:

\begin{equation}
\dot{d}_\mathrm{cc}(t)=\sum_{t_\mathrm{b}}\Gamma(t_\mathrm{b})\delta t_\mathrm{b} \overline{v}(t,t_\mathrm{b})=\int^t \Gamma(t_\mathrm{b}) \overline{v}(t,t_\mathrm{b}) \rmd t_\mathrm{b}.\label{eq:current_br}\end{equation}
Here we see that $\dot{d}_\mathrm{cc}(t)$ is the driving term for the Brunel harmonics \cite{Brunel:90} and would correspond to the current density in the quasi-classical model of Kim et al. \cite{ISI:000259649700010} if multiplied by the atomic density in the generating medium.
The emitted electromagnetic field is proportional to the dipole acceleration
\begin{equation}
\ddot{d_\mathrm{cc}}(t)=\Gamma\left(t\right) \overline{v}\left(t_\mathrm{b}=t;t\right) + \int^{t} \Gamma\left(t_\mathrm{b}\right)\dot{\overline{v}}\left(t_\mathrm{b},t\right) \rmd t_\mathrm{b}.
\label{eq:d_accel_br}
\end{equation}
%
The first term of this equation is proportional to the small initial
velocity after tunneling: $\overline{v}\left(t_\mathrm{b}=t;t\right)\approx0$, which
is neglected in the usual three-step picture, and the second one is the force $F\left[t,r(t,t_\mathrm{b})\right]=\dot{\overline{v}}\left(t_\mathrm{b},t\right)$ acting at time
$t$ on the wavepacket at the position $r(t,t_\mathrm{b})$ (center
of mass of the wavepacket). The force is given mainly by the laser
field and the Coulomb potential of the ion $F\left[t,r(t,t_\mathrm{b})\right]=F_\mathrm{las}(t) + F_\mathrm{Coul}\left[r(t,t_\mathrm{b})\right]$.
In practice, the laser field usually dominates for electrons ionized
in the strong-field regime, so the dipole acceleration is simply given by:
\begin{equation}
\ddot{d_\mathrm{cc}}(t)=F_\mathrm{las}(t)\int^{t} \Gamma\left(t_\mathrm{b}\right) \rmd t_\mathrm{b}.  \label{eq:emission_accel}
\end{equation}
The emission is simply proportional to the driving laser field
multiplied by the ionization steps $\int^{t} \Gamma\left(t_\mathrm{b}\right)\rmd t_\mathrm{b}$.
The spectrum of the emission  is obtained
by Fourier transforming Eq. (\ref{eq:emission_accel}):
\begin{equation}
	\ddot{d_\mathrm{cc}}(\omega) = F_\mathrm{las}(\omega)\star\mathcal{F}\left[ \int^{t} \Gamma\left(t_\mathrm{b}\right) \rmd t_\mathrm{b} \right](\omega), \label{eq:d_accel_spec}
\end{equation}
which results in the convolution of the laser field spectrum and the spectrum of the ionization steps.

\begin{figure}
\begin{centering}
\includegraphics[width=\textwidth]{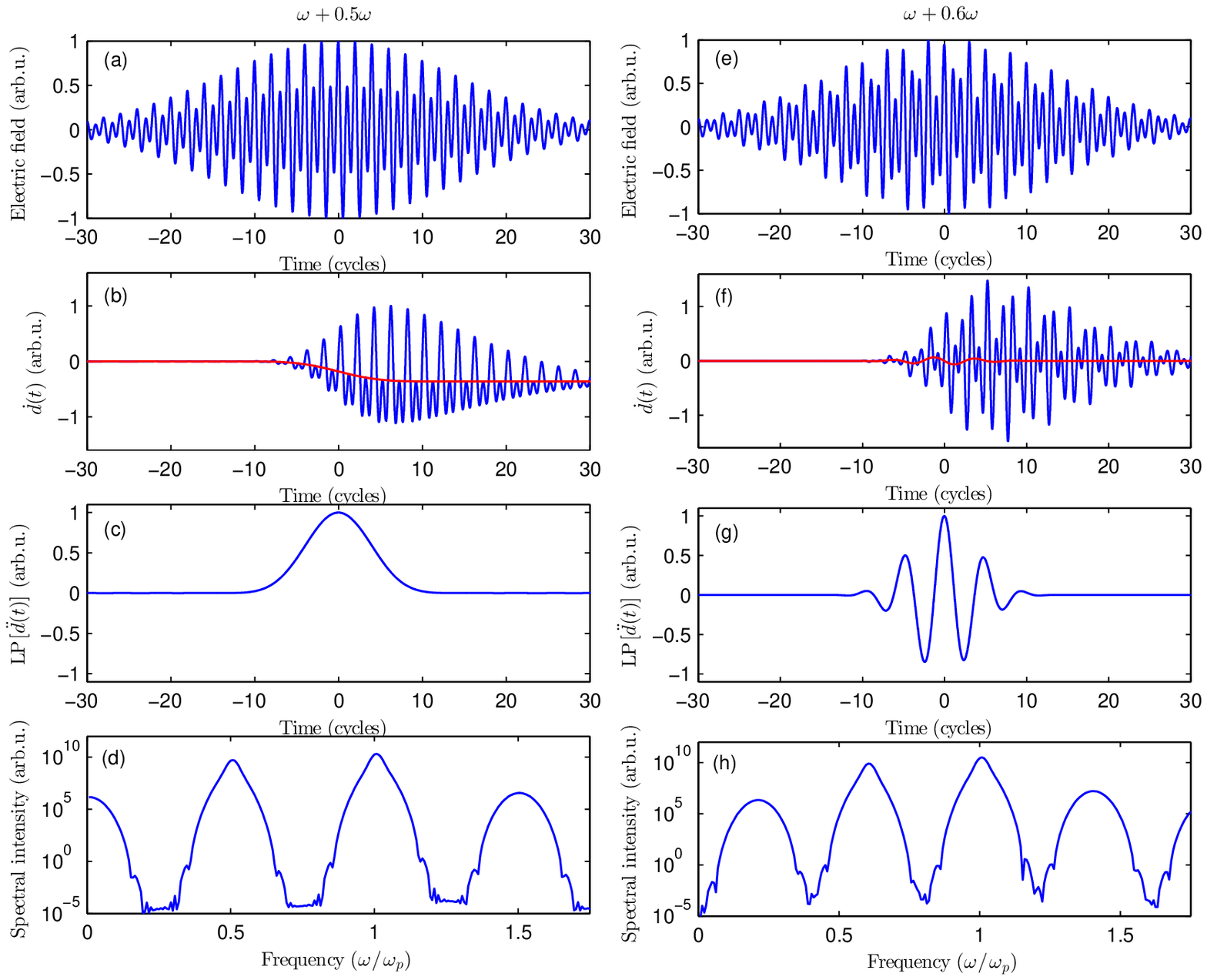}
\par\end{centering}

\caption{Simulation of the induced current, using Eqs.~\ref{eq:current_br}--\ref{eq:d_accel_spec} with $\Gamma(t_\mathrm{b})=\exp[-2(2 I_\mathrm{p})^{3/2} /(3 \vert F_\mathrm{las}(t_\mathrm{b})\vert)]$ and the classical expression $\overline{v}(t,t_\mathrm{b})=-\int_{t_\mathrm{b}}^t  F_\mathrm{las}(t') \mathrm{d}t'$, for two-colour fields with two frequency ratios: the left and right-hand sides correspond to the commensurate and incommensurate frequency case, respectively. Shown are the considered driving electric fields (first row), the dipole velocity (Eq. \ref{eq:current_br}), corresponding to the induced current (second row, blue line; the red line shows a low-pass filtered version), the low-pass filtered (retaining only the THz-sideband) dipole acceleration (Eq. \ref{eq:emission_accel}), corresponding to the emitted THz-field (third row), and the spectral intensity of the emitted radiation (square of Eq. \ref{eq:d_accel_spec}) (fourth row). The considered ionization potential is that of nitrogen molecules, $ I_\mathrm{p}=15.7\:$eV. The $\omega_\mathrm{p}$-component of the driving field has a peak field strength of $F_\mathrm{max}=27\:$GV/m (corresponding to $1\times10^{14}\:$W~cm$^{-2}$), and the $\omega_\mathrm{s}$-component has $F_\mathrm{max}=13.5\:$GV/m and a relative phase of $\pi/4$.
\label{fig:Simulation-of-current}}

\end{figure}

Now we apply the formalism derived above to the calculation of the emission spectrum in the case of a two-colour mixing scheme, which allows generation of even harmonics, including the THz sideband. Simulating our experimental conditions, the field in the calculation is composed of a strong fundamental wave at the frequency $\omega_\mathrm{p}$ and a weaker signal wave at around half the frequency of the fundamental wave $E(t)=E_{p}(t)e^{i\omega_\mathrm{p}t}+E_{s}(t)e^{i\omega_\mathrm{s}t}$, where $E_{p}(t)$ and $E_{s}(t)$ are the envelopes of the pump and signal waves respectively. The direction in which the electron is emitted is given by the sign of the velocity $v_{d}(t_\mathrm{b})=\overline{v}(t\rightarrow\infty,t_\mathrm{b})$ at which the electron drifts after the laser pulse is over. Fig.~\ref{fig:Simulation-of-current}~b) illustrates the asymmetric case of the two colour field where all of the electrons ultimately fly in one direction. The build-up of this directional electron current is responsible for the emission of the low-frequency-THz waves (see Fig.~\ref{fig:Simulation-of-current}~c,d).

Tuning of the central frequency occurs when two incommensurate frequencies are mixed. This is illustrated in the right column of Fig.~\ref{fig:Simulation-of-current} for a $0.6\omega + \omega$--combination, and compared to the commensurate case ($0.5\omega + \omega$) in the left column. While in the latter case, the growth of a directional net-current (Fig.~\ref{fig:Simulation-of-current}b) leads to the emission of a low-THz-frequency field (Fig.~\ref{fig:Simulation-of-current}c,d), the modulation of the current direction resulting from the detuning of the signal-frequency by $\Delta\omega$ leads to a $2\Delta\omega$-shift of the center of the emitted THz-spectrum (Fig.~\ref{fig:Simulation-of-current}g,h). The shift of lowest order sideband frequency is then compared to the experimentally measured THz-emission frequency. Higher order sidebands are beyond the EO sampling detection range and were not measured in this experiment.

\begin{figure}
\begin{centering}
\includegraphics[width=.6\textwidth]{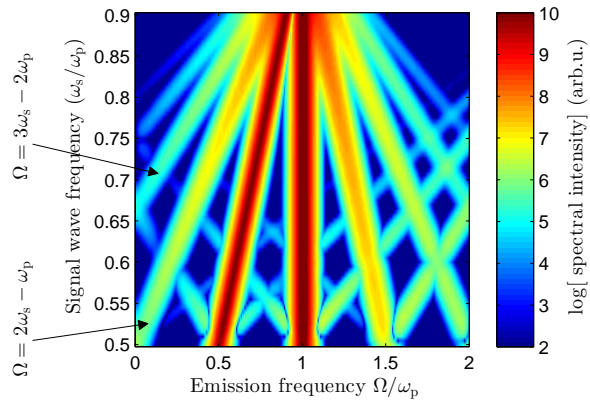}
\par\end{centering}

\caption{Simulated power spectrum of the emission (square of Eq. \ref{eq:d_accel_spec}) as a function of the frequency $\omega_\mathrm{s}$ of the tunable signal wave, for the same conditions as in figure~\ref{fig:Simulation-of-current}.
\label{fig:Simulation-of-THz}}

\end{figure}

This tunability is further analysed in Fig.~\ref{fig:Simulation-of-THz}~b). The THz-emission appears as a zeroth
order sideband alongside higher-order sidebands \cite{PhysRevLett.104.163904}.
The THz sideband starts close to zero frequency at degeneracy ($\omega_\mathrm{s}=0.5\omega_\mathrm{p}$) and its frequency increases as the frequency of the signal wave is
tuned. This effect can be interpreted as a temporal phase modulation, shifting the THz sideband towards higher frequencies as illustrated in the inset of the Fig.~\ref{fig:tuning_scheme}a).
Although the described THz emission mechanism is a phenomenon of intrinsically non-perturbative nonlinear optics, the resultant THz frequency tunability
follows the law $\Omega=2\omega_\mathrm{s}-\omega_\mathrm{p}$ (Fig.~\ref{fig:Simulation-of-THz}b), as is also the case in the perturbative four-wave mixing picture,  illustrated in Fig~\ref{fig:tuning_scheme}c) . We also predict the feasibility of generating weaker sidebands corresponding to 6, 8, etc. photon processes. This process clearly enables a tunability through the complete THz and mid-IR spectral regions~\cite{Thomson2010THzwhitelight}.

\section{Experimental implementation}
\begin{figure}
\begin{centering}
\includegraphics[width=.9\textwidth]{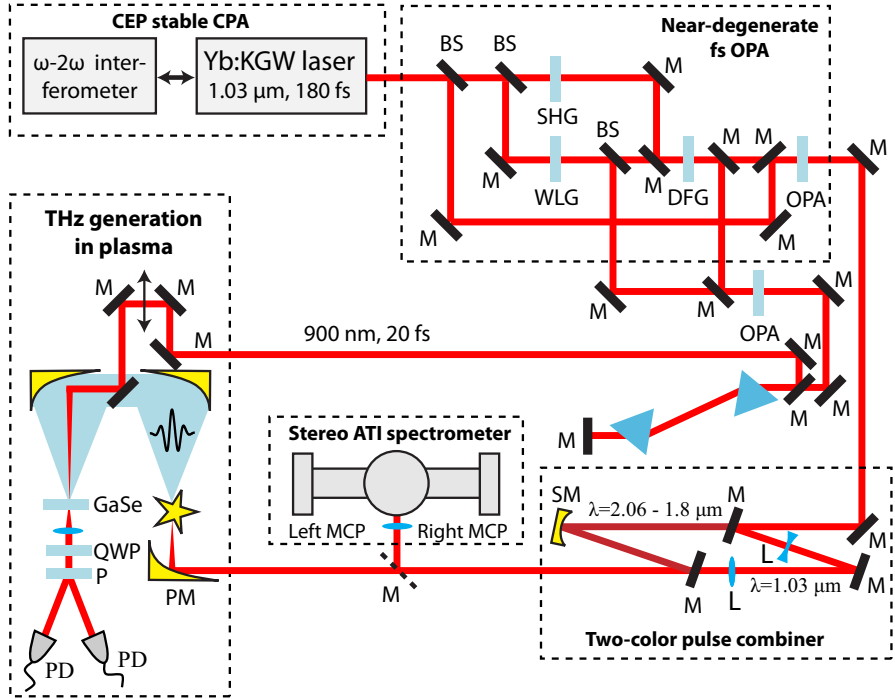}
\par\end{centering}

\caption{Scheme of the experimental setup. THz emission was detected via
electrooptic (EO) sampling setup based on GaSe crystal. The probe
pulses were generated in an additional OPA and compressed down to
20 fs in a prism compressor.\label{fig:exper-scheme}}

\end{figure}

An Ytterbium-based laser amplifier and a near-degenerate OPA were used to generate two-colour pulses that were used for plasma excitation (Fig. \ref{fig:exper-scheme}). The Yb:KGW regenerative amplifier is actively CEP stabilized~\cite{pharoscep} and is used to pump a two-stage near-degenerate OPA. First, part of the laser energy is split into three parts. One part is used to generate a broadband super-continuum by focusing it in a sapphire plate. The part of the so-generated spectrum around 680--720~nm  is used as a seed for the first OPA stage. The second part of the beam is frequency-doubled and serves as a pump for the first OPA stage. The resulting difference-frequency (idler) wave from the first stage (1800~nm to 2100~nm) is then amplified as a signal wave in the second OPA stage, pumped by the third part of the fundamental laser beam. The resulting CEP of the OPA output is thus equal to the pump laser pulse CEP phase \cite{PhysRevLett.88.133901}, up to a constant: $\varphi_\mathrm{s}=\varphi_ \mathrm{p}+\mathrm{const.}$. 

At the output of the OPA, the pump wave at $\lambda_\mathrm{p}=1030\:$nm and the signal wave at $\lambda_\mathrm{s}=$1800--2100~nm are separated using dichroic mirrors and then re-combined after adjusting the divergence of the beams and the relative delay. The polarisations of each of the two colour-component pulses are linear and set parallel to each other. The phase delay between the two constituent colour waves is $\tau_\phi=\varphi_ \mathrm{p}(\lambda_\mathrm{s}-\lambda_\mathrm{p})(2\pi c)^{-1} + \mathrm{const.}$, and can thus be controlled via the fundamental laser CEP $\varphi_ \mathrm{p}$.

\subsection{Observation of directional electron current}

We verified that the two-colour pulses used for THz generation induce a directional electron drift by measuring the photoelectron spectra produced via above threshold ionization (ATI) with a stereo-ATI spectrometer \cite{ISI:000172029100040,Wittmann2009}. The electron spectrometer consists of a small xenon-filled gas cell ($<1\:$mbar) placed in a vacuum chamber with two identical time-of-flight (TOF) electron spectrometers oriented opposite each other in order to measure the photoelectron spectrum emitted in the opposite directions. The generated photo-electrons pass through small holes in the gass cell into the TOF arms. Each TOF arm consists of a field-free propagation tube ($\approx50$~cm) and a micro-channel plate (MCP) electron detector. The linear polarisation of each of the two-colour driving wave is in the direction of the TOF detectors. The signals from the two TOF detectors are then acquired with multi-scaler card and processed using a computer.

\begin{figure}
\begin{centering}
\includegraphics[width=.8\textwidth]{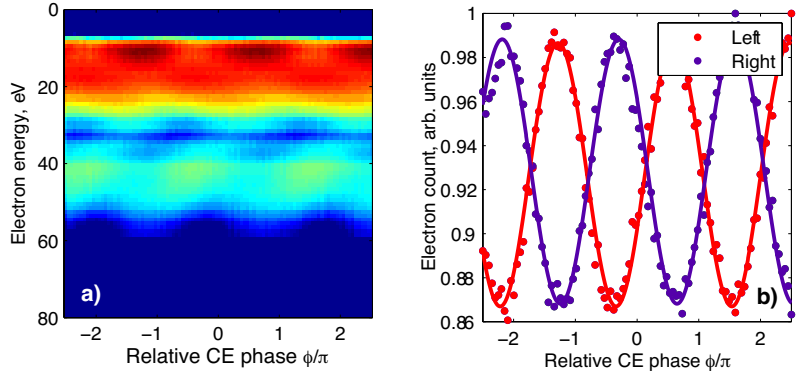}
\par\end{centering}
\caption{Stereo-ATI electron spectrometry measurement results obtained with two commensurate-frequency pulses: $\lambda_\mathrm{p}=1.03\:\mu$m and $\lambda_\mathrm{s}=2.06\:\mu$m. (a) Electron energy spectrum (log scale) measured on one side of the stereo-ATI spectrometer as function of the fundamental laser CEP. (b) Total electron count in left and right directions of the stereo-ATI spectrometer obtained by integrating the electron spectrum of the left and right detectors.
\label{fig:Experimental-results-ati}}
\end{figure}

The ATI spectrum in one direction measured for different CEP phases of the laser pulse is shown in Fig.~\ref{fig:Experimental-results-ati}~a).This measurement is done with the OPA tuned at degeneracy ($\lambda_\mathrm{s}=2060\:$nm) so that the maximum directional drift is produced, as illustrated in the left column of figure \ref{fig:Simulation-of-current}. The asymmetry of the low-energy direct electrons is opposite to that of high-energy rescattered electrons (not considered in figure \ref{fig:Simulation-of-current}) and partially reduces the asymmetry of the total electron yield. However, because the number of direct electrons is much higher than of the rescattered ones, the total yield still exhibits a modulation depth higher than 15\%, as shown Fig.~\ref{fig:Experimental-results-ati}~b). These findings prove that in our experimental conditions, the induced electron current is highly directional and we are working in the tunnelling regime, as assumed for the theoretical model.

\subsection{CEP-locked, tunable THz emission}
We now demonstrate the phase-locked temporal modulation of this induced electron current to achieve tunability of resulting CEP-locked THz emission.

\begin{figure}
\begin{centering}
\includegraphics[width=\textwidth]{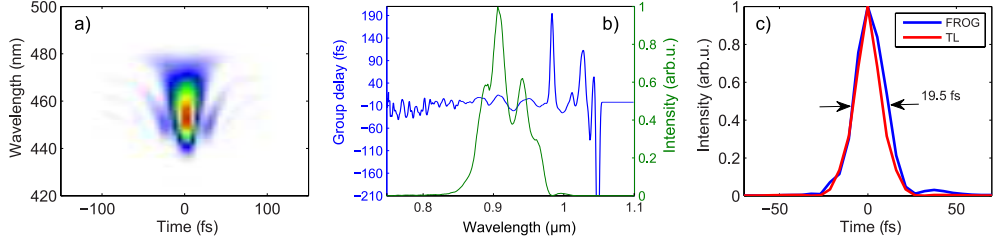}
\par\end{centering}

\caption{Temporal characterisation of EO sampling probe pulse using SHG FROG. a) measured FROG trace, b) reconstructed spectrum and group delay of the pulse c) measured temporal profile and calculated Fourier limit.
\label{fig:probe_pulse_Frog}}

\end{figure}

For THz-generation, the combined two-colour driving waveform is focused using an $f=7.5\:$cm parabolic mirror into ambient air. The field of the generated THz radiation is detected using electro-optic (EO) sampling. In general, the bandwidth $f_\mathrm{max}$ in the EO sampling detection scheme depends on the probe pulse duration $f_{max}\approx 1/(2\tau_\mathrm{probe})$ and it is the main limitation in our setup. The probe pulse of $\tau_\mathrm{probe}\approx 20$~fs duration was generated by amplifying the spectral portion at around 900 nm of the white-light super continuum in a separate OPA stage and then compressing the pulse in a prism compressor near transform limit (see figure \ref{fig:exper-scheme}). The probe pulse was characterised using SHG Frequency Resolved Optical Gating (FROG) with the result shown in Fig.~\ref{fig:probe_pulse_Frog}. 

The phase of the THz wave according to the frequency dependence
is $\varphi_\mathrm{THz}=2\varphi_\mathrm{s}-\varphi_\mathrm{p}$. Therefore, the degenerate scheme
where fundamental and frequency-doubled pulses are mixed (with CEPs $\varphi_{2\omega} = 2\varphi_{\omega}+\mathrm{const.}$) provides
a passive stabilization of the THz-pulse's CEP, ($\varphi_{THz}=2\varphi_{\omega}-\varphi_{2\omega}=\mathrm{const}$),
similar to difference-frequency generation based on a second-order nonlinearity
($\chi^{2}$) \cite{PhysRevLett.88.133901}. Since in our case,  $\varphi_{s} = \varphi_{p}+\mathrm{const.}$,
the CEP of the THz-pulse is the same as that of the fundamental laser pulse: $\varphi_{THz}=\varphi_{p}+\mathrm{const}$.
Because of the field-sensitive EO sampling of the THz emission, active stabilization of the laser CEP is thus mandatory. In other words, for non-CEP-stabilized driving pulses, the THz emission is still emitted, but their CEP is random from pulse to pulse.

\begin{figure}
\begin{centering}
\includegraphics[width=.9\textwidth]{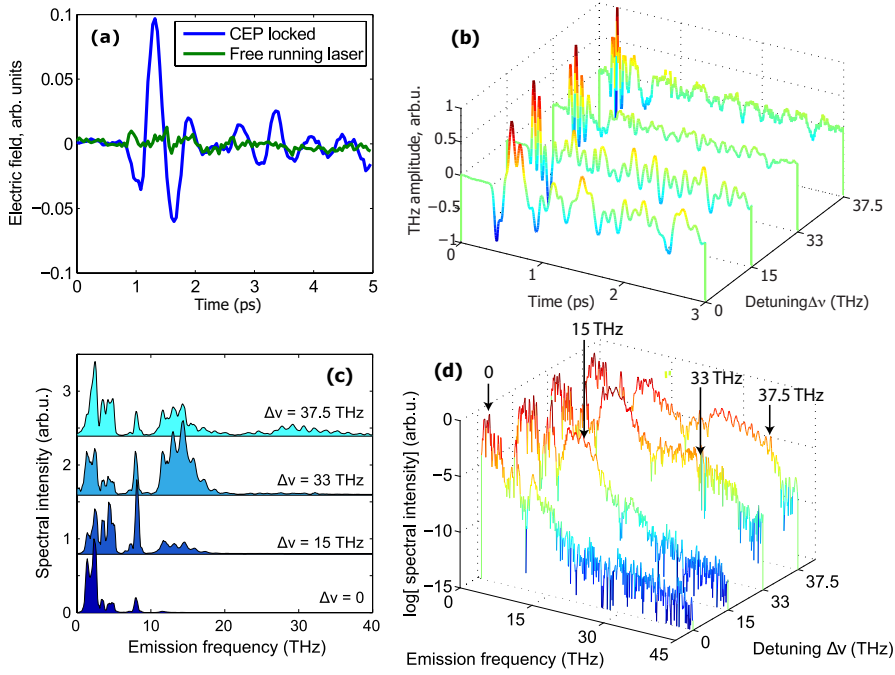}
\par\end{centering}

\caption{Experimental THz-generation results with $\omega_\mathrm{p}+\omega_\mathrm{s}$ two-colour driving fields ($\omega_\mathrm{s}=(\omega_\mathrm{p}+\Delta\nu)/2$), for various detunings $\Delta\nu$, created by combining the fundamental wave ($\lambda_\mathrm{p}=1.03\:\mathrm{\mu m}$, 250~$\mathrm{\mu J}$) with the tunable signal wave ($\lambda_\mathrm{s}=[2.06, 1.96, 1.84, 1.80]\:\mathrm{\mu m}$, 20 $\mathrm{\mu J}$). (a) Electric field of the THz transient measured using EO sampling in case of locked CEP or free running laser, for zero detuning, $\Delta\nu=0$, i.e. exactly commensurate frequency ratio. (b) Temporal traces of the EO sampling signal for different detunings $\Delta\nu$, (c) THz emission spectra of plasma calculated from the EO signal, represented on linear scale. (d) The same spectra as in (c), represented on logarithmic scale. Each spectrum and temporal trace was normalised to its maximum. The arrows in panel (d) show the theoretically predicted central frequency of the emission.
\label{fig:Experimental-results-thz}}
\end{figure}

This is why, as shown in Fig.~\ref{fig:Experimental-results-thz}a), without active CEP stabilization of the laser pulse, the emitted THz waveforms for each laser pulse are averaged out in our EO sampling. By locking the laser CEP, we create reproducible THz fields that let us study the two-colour-driven plasma dynamics.

The detected THz emission spectra as function of the signal wavelength $\lambda_\mathrm{s}\leq2\lambda_\mathrm{p}$ are shown in Fig.~\ref{fig:Experimental-results-thz}~b).
Despite the rather narrow bandwidth of our 180-fs pulses that were used to excite the plasma, mid-IR emission frequencies up to $\approx40\:$THz were measured as the result of detuning the OPA from degeneracy. The frequency of the THz-sideband shifts towards higher frequencies due to temporal current modulation when the OPA is detuned from degeneracy, as illustrated by the numerical simulations shown in figure~\ref{fig:Simulation-of-current}. We can detect only the lowest THz-sideband in the spectrum due to the limited bandwidth of the THz detection apparatus. At higher frequencies, the probe pulse duration is significantly longer than the half-cycle of the field to be measured. The decreasing sensitivity of the THz detection setup at higher frequencies also contributes to the broadening of the spectrum for higher frequency detuning. The experimentally obtained frequency dependence agrees well with the semi-classical transient electron-current model simulations in figures \ref{fig:Simulation-of-current} and \ref{fig:Simulation-of-THz}: we observe a shift by $\approx\Delta\nu$ for a signal wavelenegth tuned to $\omega_\mathrm{s}=(\omega_\mathrm{p}+\Delta\nu)/2$.

\section{Conclusion}
The tunability and CEP-stability of the THz emission from a laser-induced plasma-spark is demonstrated by mixing two incommensurate CEP-stable optical driving fields, forming a parametric waveform synthesizer producing shot-to-shot reproducible strong-field multi-colour waveforms. The measured frequency dependence of the THz emission is consistent with the transient electron current model we derived based on continuum-continuum transition matrix elements.

\section*{Acknowledgements}
These studies were supported by the ERC project CyFi 280202 and Austrian Science Fund (FWF) via the SFB NextLite F4903-N23. S. Haessler acknowledges support by the EU-FP7-IEF MUSCULAR and the by the Austrian Science fund (FWF) through grant M1260-N16.

\section*{Appendix: Stationary phase analysis}

An evolution of a continuum electron wavepacket can in general be described by the propagator (\ref{eq:wavepacket_propagation}).
For convenience we will replace the integral by a sum of small wavepackets
corresponding to different birth times $t_\mathrm{b}$:

\begin{equation}
\psi_{t_\mathrm{b}}(r,t,t_\mathrm{b})=-i\hat{U}(t,t_\mathrm{b})\hat{V}(t_\mathrm{b})\hat{U}_{0}(t_\mathrm{b},0)\Psi_{0}\delta t_\mathrm{b}\end{equation}
where $\delta t_\mathrm{b}$ is associated with the width of the saddle point
region around $t_\mathrm{b}$ that satisfies the stationary phase condition:
\begin{equation}
\vec{r}=\vec{r}_{0}+\int_{t_\mathrm{b}}^{t}\vec{v}\left(t^{\prime\prime},t\right)dt^{\prime\prime}.
\label{eq:stationary_phase_condition}
\end{equation}
Here $\vec{r}_{0}$ is within the size of the initial state $\Psi_{0}$
and is therefore negligible compared to the oscillation of the electron
in the strong laser field described by the second term.
This condition is valid beyond the SFA \cite{PhysRevA.77.033407}.
By coherently adding up the small wavepackets of different birth times
we get the wavepacket of the electron in the continuum,  Eq. (\ref{eq:sum_of_wavepackets}).
The evolution of $\psi_{t_\mathrm{b}}(t)$ is governed by both the ionic potential
and the laser field. To calculate the emission, the dipole velocity 
$\dot{d}_\mathrm{cc}(t) = \left\langle \Psi_\mathrm{c}\left|\hat{p}\right|\Psi_\mathrm{c}\right\rangle = \sum_{t_\mathrm{b1},t{}_\mathrm{b2}}\left\langle \psi_{t_\mathrm{b2}}\left|\hat{p}\right|\psi_{t_\mathrm{b1}}\right\rangle$ has to be calculated. In case of a strong
laser field, the greatest contribution to the the momentum operator
$\hat{p}$ is given by the derivative of the action $S$ which is a rapidly
oscillating function:

\begin{equation}
i\nabla\psi_{t_\mathrm{b}}(t)\approx-\sqrt{\rho(t,t_\mathrm{b})}\nabla S(t,t_\mathrm{b})e^{iS(t,t_\mathrm{b})}=-\frac{\partial S}{\partial r}\psi_{t_\mathrm{b}}(t)\end{equation}
Then the dipole velocity can be written as:
\begin{eqnarray}
\dot{d}_\mathrm{cc}(t) 
&=\sum_{t_\mathrm{b1},t_\mathrm{b2}}\left\langle \psi_{t_\mathrm{b1}}\left|\frac{\partial S\left(t,t_\mathrm{b2}\right)}{\partial r}\right|\psi_{t_\mathrm{b2}}\right\rangle \\
&=\sum_{t_\mathrm{b1},t_\mathrm{b2}}\int dr\sqrt{\rho_{1}}\sqrt{\rho_{2}}e^{i\left[S(r,t,t_\mathrm{b2})-S(r,t,t_\mathrm{b1})\right]}v(r,t,t_\mathrm{b1},t_\mathrm{b2})
\end{eqnarray}
where $v=\partial S/\partial r$ is the velocity of the electron
at point $r$, $\rho_{1}\equiv\rho(t,t_\mathrm{b1},r)$ and $\rho_{2}\equiv\rho(t,t_\mathrm{b2},r)$
are the densities of the two wavepackets. The non-negligible contribution
to the sum is given by the diagonal terms $t_\mathrm{b1}=t_\mathrm{b2}$ that satisfy
the stationary phase condition. By taking into account only these
diagonal terms and calculating the acceleration of the dipole, we
finally arrive at equation (\ref{eq:emission_accel}) which is the
same as the one used in photocurrent calculations.

\section*{References}
\bibliography{refdb8}

\end{document}